\documentclass[a4paper,11pt]{article}
\usepackage{jinstpub} %
\usepackage{lineno}

\usepackage{caption}
\DeclareUnicodeCharacter{2212}{-}
\usepackage{lipsum} 
\usepackage{siunitx}
\usepackage{amsmath}
\usepackage{xcolor}
\usepackage{subfig}
\usepackage{comment}
\usepackage{isotope}
\usepackage{ragged2e}

\title{\boldmath Cryogenic SiPMs for the Optical Readout of DarkSide-20k}

\author{G. Matteucci on behalf of the DarkSide-20k Collaboration}
\affiliation{Physics Department, Università degli Studi “Federico II” di Napoli,\\Via Cintia 21, Napoli 80126, Italy}
\affiliation{Istituto Nazionale di Fisica Nucleare, Sezione di Napoli,\\Via Cintia 21, Napoli 80126, Italy}

\emailAdd{giuseppe.matteucci@infn.it}

\abstract{Silicon photomultipliers (SiPM) have gained significant traction as an alternative technology to the well-established photomultiplier tube (PMT), with numerous high-sensitivity experiments adopting them either complementarily or as a replacement for PMTs. SiPMs are an ideal match for low-background cryogenic applications, such as massive noble liquid experiments for dark matter direct detection, due to (i) the significant reduction of dark noise in cold environments, (ii) relatively low radioactive content, and (iii) scalable industrial production. For these reasons, the Global Argon Dark Matter Collaboration has committed to this technology for DarkSide-20k, its next experiment for the direct search of WIMP Dark Matter, currently in construction at LNGS Hall C. The development of a large-area cryogenic SiPM-based photon counter has culminated in the Photon Detector Unit (PDU), a compact photosensor measuring $20\times20$ cm$^2$ with $100$ cm$^2$ active surface per channel, based on SiPM technology from Fondazione Bruno Kessler and incorporating custom front-end electronics suited for cryogenics. More than 600 PDUs are being produced and tested in various collaboration facilities to construct the two $~10.5$ m$^2$ optical planes of the massive two-phase argon time projection chamber of DarkSide-20k and the optical readout of its veto system.}


\begin{document}
\maketitle
\flushbottom

\section{The DarkSide-20k Experiment}
\justifying

Dark matter (DM) constitutes about $85\%$ of the matter content of the Universe according to the $\Lambda$CDM model \cite{Planck:2018nkj,Planck:2018vyg}, and yet, its nature remains unknown.
One promising candidate is the Weakly Interacting Massive Particle (WIMP) \cite{Roszkowski:2004jc}, which could be revealed in terrestrial detectors through coherent elastic scattering on nuclei, depositing an energy in the $(1\text{--}100)\,\mathrm{keV}$ range. The WIMP-nucleon interaction cross-section has already been constrained by many orders of magnitude for a wide range of masses in the last decades \cite{Billard:2021uyg}, and future experiments must scale up their detectors while preserving a background level low enough to enable a potential discovery.
For background reduction, liquid argon (LAr) detectors have proven to be particularly effective thanks to the extremely powerful pulse-shape discrimination techniques which exploit the peculiar scintillation properties of argon \cite{Matteucci:2024foy}. The Global Argon Dark Matter Collaboration (GADMC) aims to advance LAr-based detectors to the so-called `neutrino fog' \cite{OHare:2021utq} sensitivity.

\begin{figure}[t]
    \centering
    \includegraphics[height=0.45\linewidth]{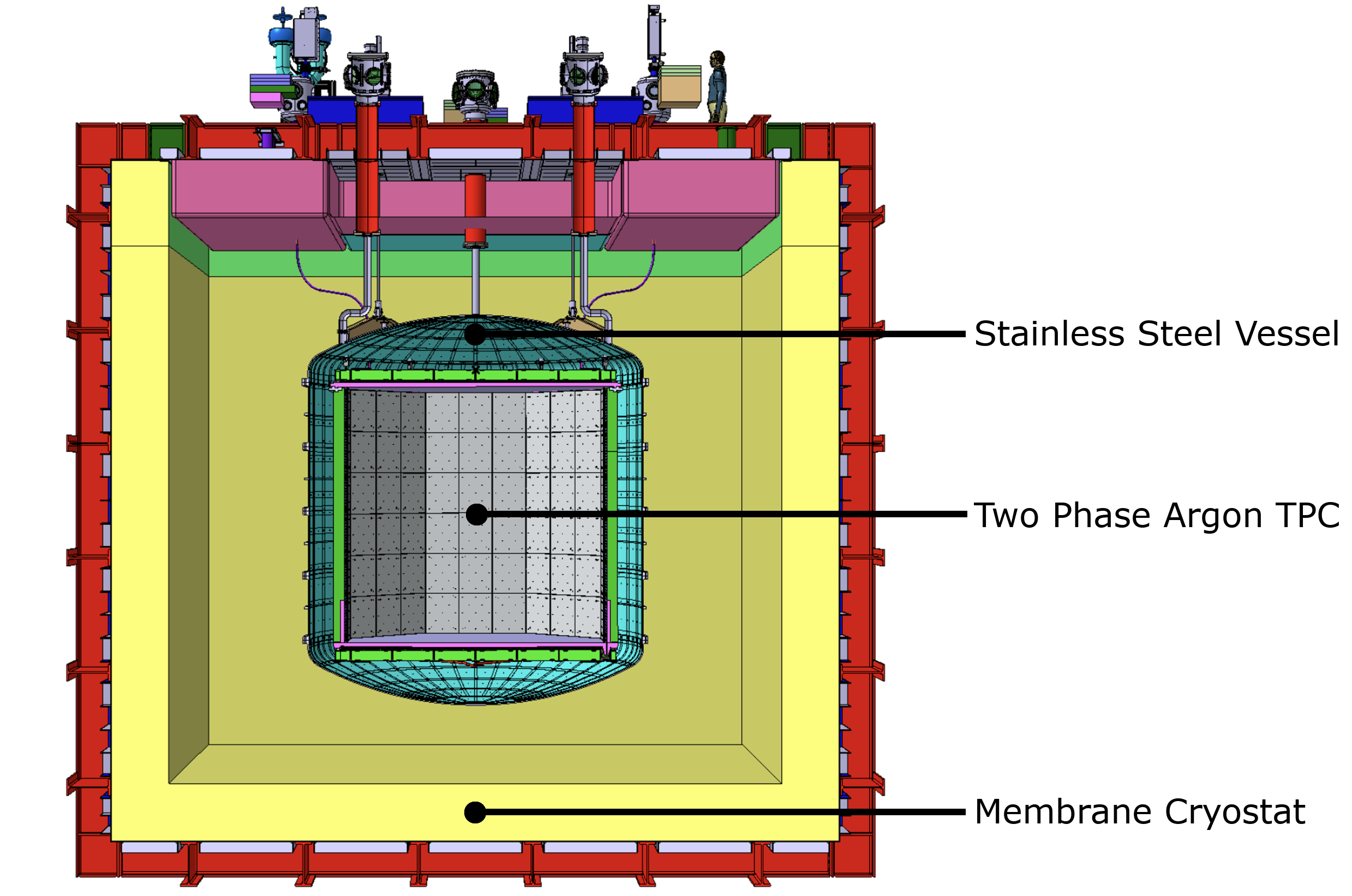}
    \caption{Section of the DarkSide-20k detector assembly. Inside the cryostat lies the inner detector, composed of the TPC and the inner veto system, which are contained within a stainless steel vessel filled with radiopure, underground-extracted argon. Atmospheric liquid argon is used instead in the main cryostat, which is also instrumented with photosensors to serve as muon veto.}
    \label{fig:ds20k_and_veto}
\end{figure}

The next-generation GADMC experiment is DarkSide-20k (DS-20k), a massive argon detector currently under construction at LNGS underground laboratories (Figure~\ref{fig:ds20k_and_veto}). The core of DS-20k is a two-phase argon time projection chamber (TPC) with a fiducial argon mass of \SI{20}{\tonne}, with maximum sensitivity in the DM mass range of \SI{1}{\GeV/c^2} to \SI{10}{\TeV/c^2}. The versatility of the two-phase argon TPC enables DS-20k to be highly competitive in lighter dark matter searches \cite{DarkSide-20k:2024yfq} and to join the search for core-collapse supernova neutrinos via coherent elastic neutrino-nucleus scattering \cite{DarkSide20k:2020ymr}. DS-20k is intended to reach an exposure of a few hundred \unit{\tonne\cdot yr} with a target instrumental background of less than \(0.1\) events in the DM region of interest.

DarkSide-20k has a nested detector structure formed by two anti-coincidence (veto) systems enveloping the central two-phase TPC. %
The expected interaction from a WIMP, a coherent elastic scattering on an argon nucleus, can be detected in the TPC via a combined measurement of scintillation light (S1 signal) and ionization electrons (S2 signal) produced in LAr by the recoiling argon nucleus.
Beyond the increased size, DarkSide-20k introduces many innovations over its predecessor DarkSide-50 \cite{DarkSide:2014llq,DarkSide:2018kuk}, with the most significant being the use of an optical readout system leveraging large-area, radiopure, cryogenic SiPM photon counters for both the TPC and the veto systems. %

The TPC of DS-20k is a regular octagonal prism with an inscribed-circle diameter and height of \SI{350}{\cm}. It is capped at the top and bottom by two transparent acrylic windows coated with a thin conductive polymer (Clevios) that serves as an electrode for both the drift and extraction fields, necessary to drive ionization electrons into a thin gas layer where they produce secondary scintillation. 
The TPC comprises, together with the inner veto, the inner detector. The two systems are contained in a stainless steel vessel filled with low-radioactivity underground-extracted argon (UAr). The use of UAr is essential to reduce the radio contamination from the \isotope[39]{Ar} isotope, abundant in atmospheric argon (AAr) ~\cite{WARP:2006nsa}. The radioactivity of UAr for DS-20k has been measured by DarkSide-50 to be 1400 times lower than that of AAr~\cite{DarkSide:2015cqb}. The URANIA project by the GADMC will procure all the necessary UAr for DS-20k from an underground CO\textsubscript{2} well in Colorado, USA. The UAr will then be shipped to the Aria facility \cite{DarkSide-20k:2021nia} in Sardinia for chemical purification~\cite{DarkSide-20k:2021nia}.%

The stainless steel vessel of the inner detector is itself placed inside a ProtoDUNE-like membrane cryostat~\cite{Montanari:2015pxz} with a volume of about \SI{600}{\cubic\meter} and filled with atmospheric liquid argon, instrumented to form the outer veto. %

\section{PDUs for Light Detection in DarkSide-20k}
Argon scintillation light is emitted in the VUV region at about \SI{128}{nm}, which is then shifted to visible wavelengths using tetraphenyl-butadiene (TPB) deposited on internal detector surfaces. The wavelength-shifted light is collected by two SiPM-based optical planes, each covering an area of $\sim10.5\,\mathrm{m^2}$.%

SiPMs were selected as the photon detection technology for DS-20k due to their inherently low radioactivity~\cite{Baudis:2018pdv}, strong performance~\cite{Renker:2006ay}, and suitability for scalable manufacturing processes. The SiPMs utilized in DS-20k were specifically developed for cryogenic operation in collaboration with Fondazione Bruno Kessler (FBK)~\cite{Gola:2019idb}.

\begin{figure}[t]
    \centering
        \subfloat[\centering]{\includegraphics[height=80pt]{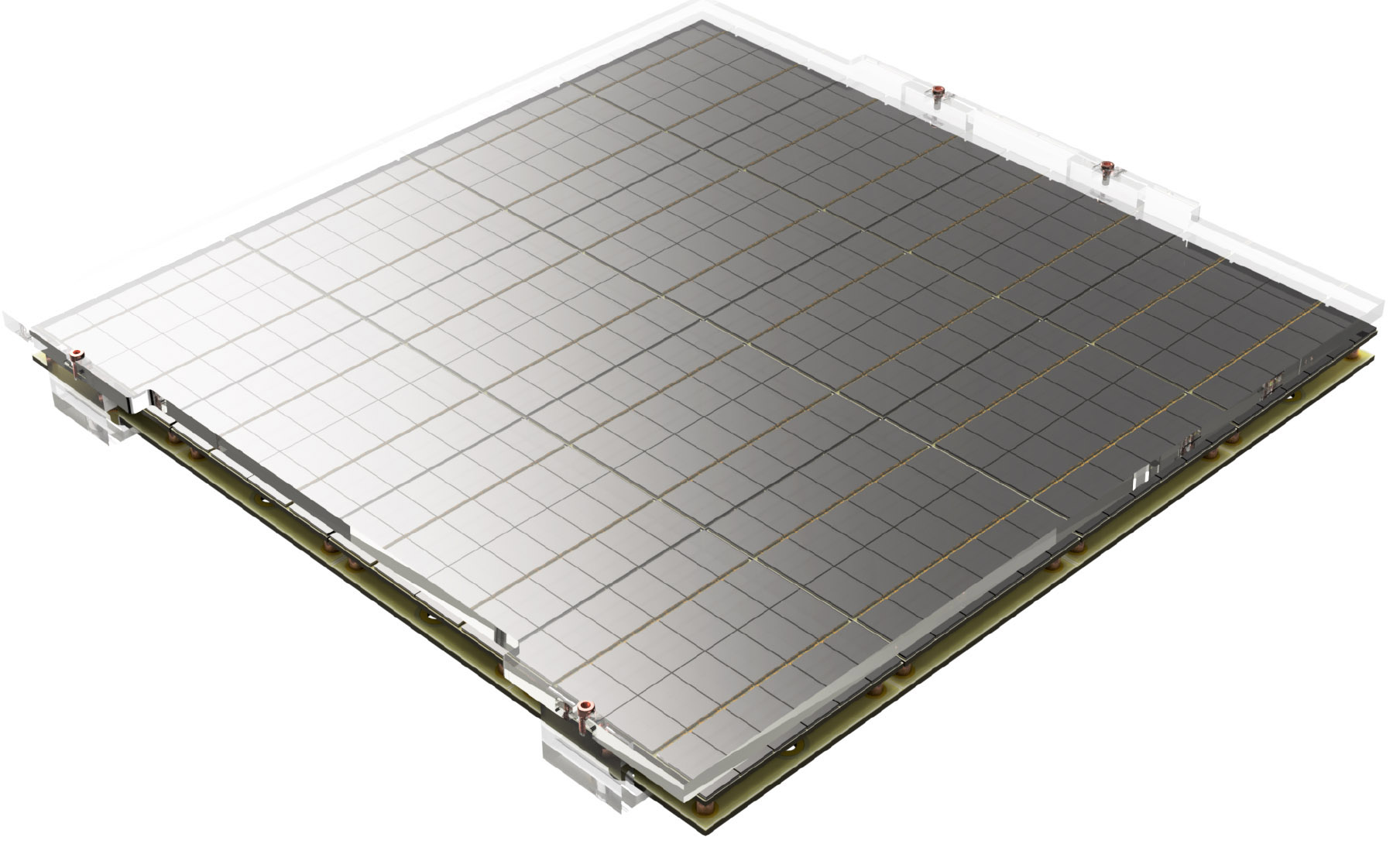}}
        \qquad
    \subfloat[\centering]{\includegraphics[height=80pt]{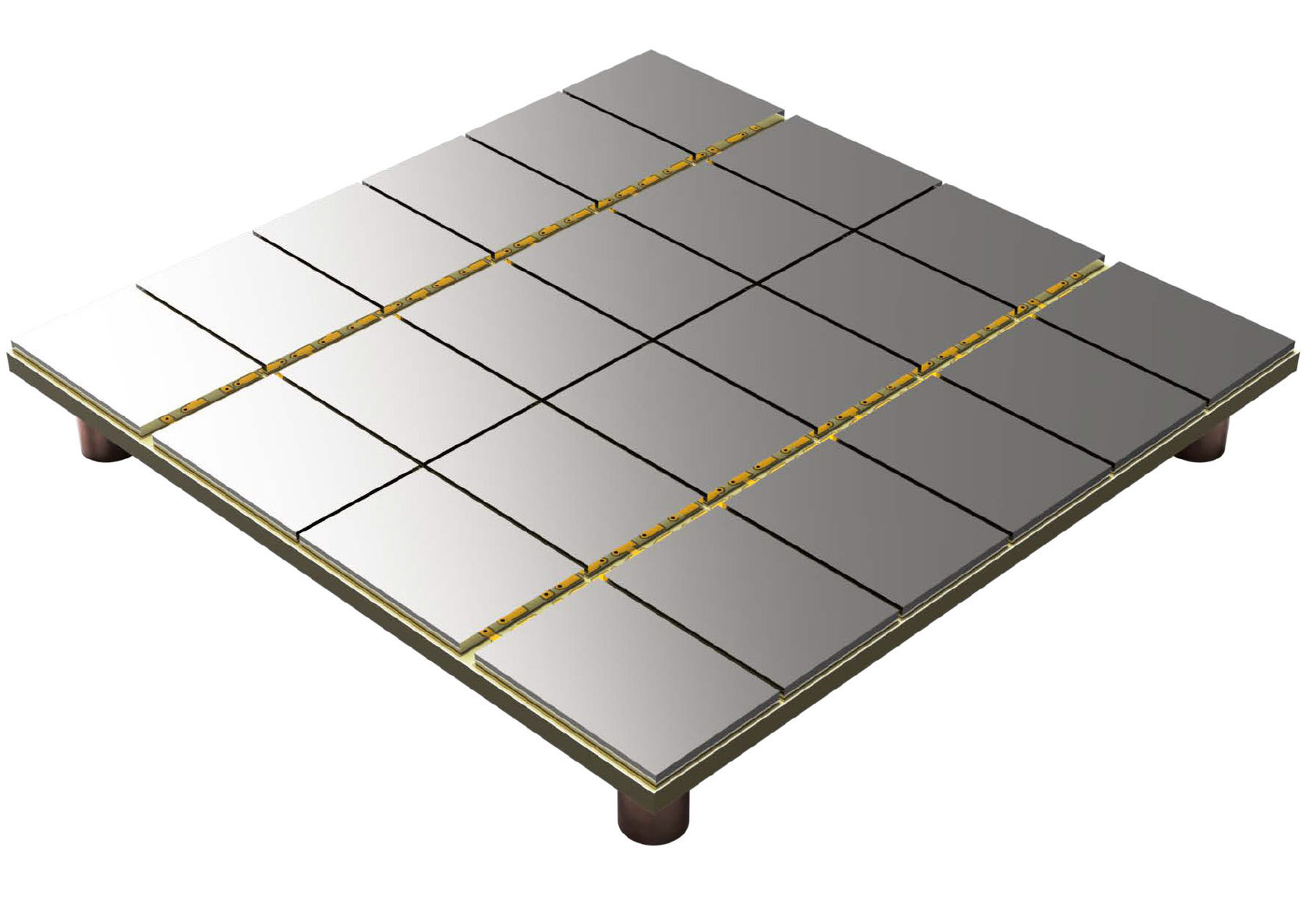}}
    \qquad
        \subfloat[\centering]{\includegraphics[height=80pt]{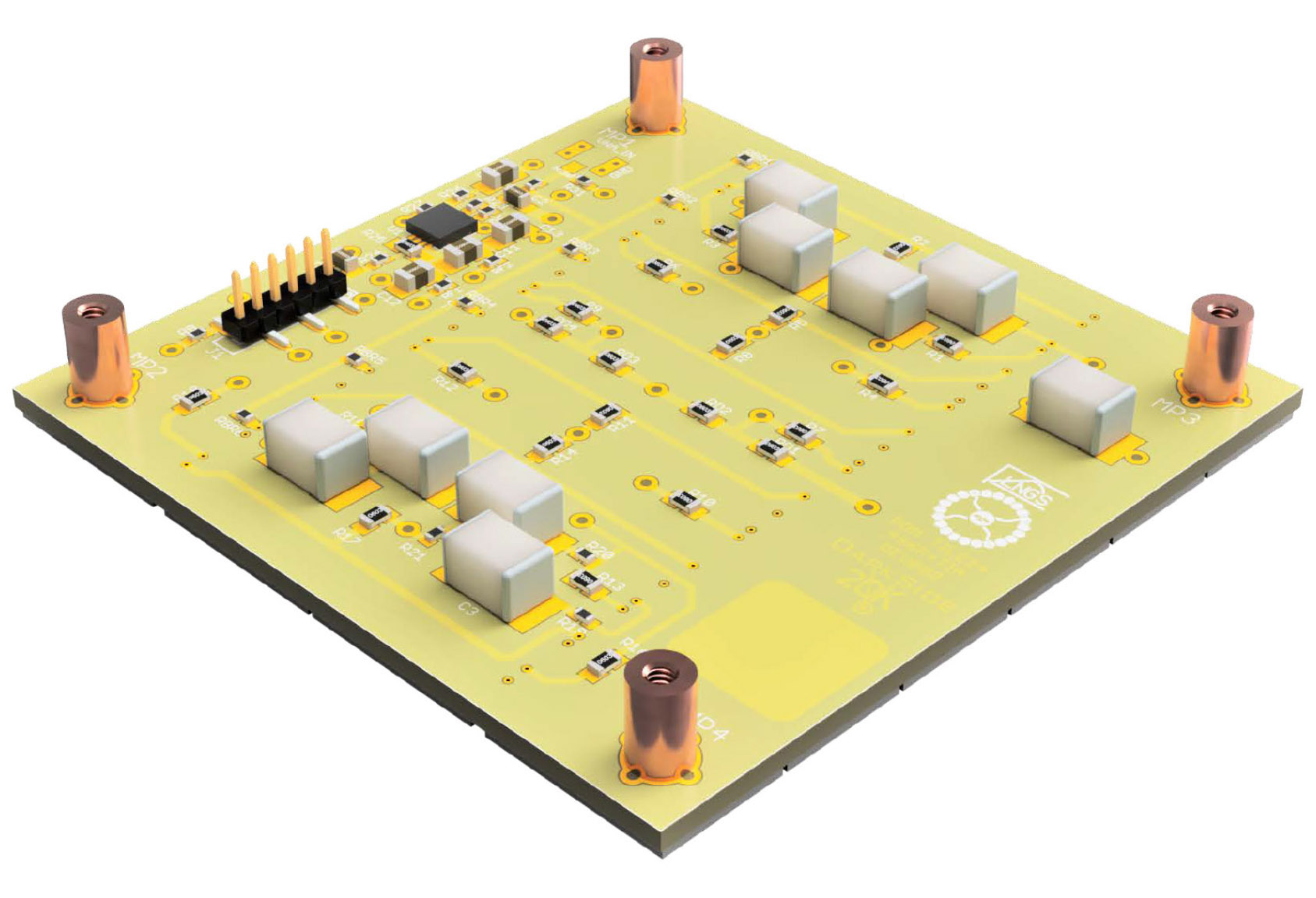}}

    \caption{\textit{(a)} Picture of the fully assembled PDU, with a total surface of $20\times20$~\unit{\cm\squared}. The transparent cover is for protection during transport and handling. \textit{(b)} Top view of a DarkSide-20k $5\times5$~\unit{\cm\squared} tile, with the 24 SiPMs visible. \textit{(c)} Bottom view of the tile, showing the front-end electronics. }
    \label{fig:pdupdm}
\end{figure}

To instrument the large-scale optical system of DS-20k, the Collaboration conducted an extensive R\&D program aimed at developing a modular and scalable design. The result of this effort is the Photon Detector Unit (PDU). The PDU is a modular, cryogenic photon detector with a total area of $20\times20\,\text{cm}^2$. Each PDU provides four analog readout channels, each channel covering $100\,\text{cm}^2$ of active SiPM area. Figure~\ref{fig:pdupdm}a shows a fully assembled PDU. 

Each readout channel consists of four \emph{tiles}, being made of 24 SiPMs die-bonded onto a custom PCB. The SiPMs within a tile are connected in a parallel-series configuration to limit the overall capacitance. A trans-impedance amplifier (or a custom ASIC for PDUs in the the veto systems) is mounted on the back side of the tile PCB (Figure~\ref{fig:pdupdm}c) and converts the signal from the SiPMs into an analog voltage signal. On the main PDU motherboard, signals from tiles are combined via active summers and then converted from single-ended to differential format. A power management section enables remote on/off control for each tile. The total power consumption at LAr temperature is less than \SI{2}{W} per PDU.

For the DarkSide-20k TPC, a total of 528 PDUs will be installed, providing 2112 readout channels. An additional 160 PDUs are distributed on the surfaces of the inner and outer veto systems to provide light readout. Monte Carlo simulations indicate that the combination of high SiPM photon detection efficiency (PDE)~\cite{Acerbi:2022xpt}, extensive coverage, and reflective walls will yield a total light collection efficiency of about $45\%$ in the TPC.

\begin{comment}
\begin{figure}
    \centering
    \includegraphics[width=0.6\linewidth]{figs/tile_on_a_motherboard.png}
    \caption{Render of a Photon Detector Unit (PDU) showing the assembly of a SiPM tile on the PDU motherboard.}
    \label{fig:tileassembly}
\end{figure}
\end{comment}

\section{Production and Testing of PDUs}
PDU production starts from SiPM wafers, which are mass-produced by LFoundry and shipped to the recently established INFN-LNGS facility Nuova Officina Assergi (NOA)~\cite{DarkSide-20k:2024usz,Consiglio:2023nkj}. NOA boasts a \SI{420}{\meter\squared} ISO~6 clean room, equipped with all the necessary machinery for tile and PDU assembly. The clean environment contributes to constraining the radio-contamination in DarkSide-20k.

From NOA, PDUs are sent to the Photosensor Test Facility (PTF)~\cite{DarkSide:2022nfm} located at the Cryogenic Laboratory of INFN Naples. A comprehensive test protocol is performed on batches of up to 16~PDUs, at both room temperature and in liquid nitrogen (LN).

Concurrently, the production and testing of Veto PDUs are being carried out by several institutions: the University of Birmingham for veto tile assembly, the University of Liverpool and STFC-Interconnect for SiPM die attach and wire-bonding, the University of Oxford for ambient temperature and cryogenic testing of tiles, the University of Manchester for veto PDU assembly and ambient temperature testing, the University of Liverpool for cryogenic testing, with additional facilities at the University of Edinburgh, the University of Lancaster, and AstroCeNT (Poland).

\section{Performance of the PDU}
\begin{figure}[t]
    \centering
        \subfloat[\centering]{\includegraphics[height=130pt]{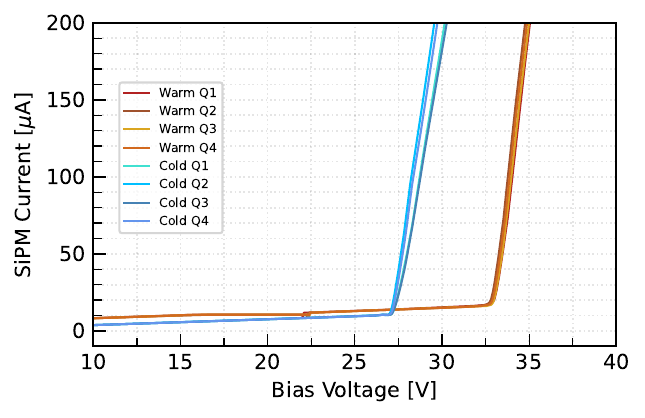}}
        \qquad
    \subfloat[\centering]{\includegraphics[height=130pt]{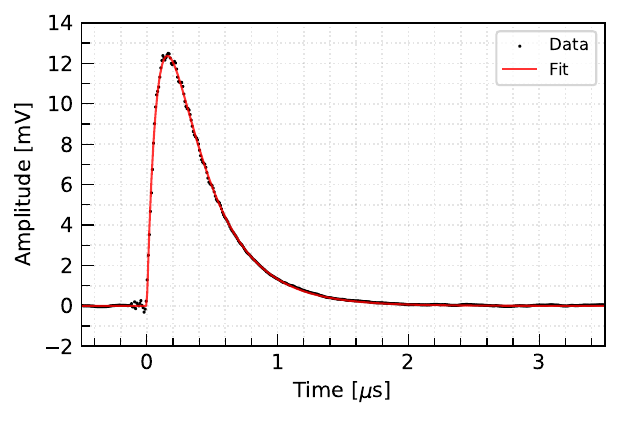}}

    \caption{\textit{(a)} The I-V characteristic of the four channels of the PDU with respect to SiPM bias voltage, at ambient temperature (labelled as \textit{warm}) and in liquid nitrogen (labelled as \textit{cold}). At \SI{77}{\K} the breakdown voltage is \SI{27.1(1)}{V}. \textit{(b)} Average waveform for the signal associated with stimulation from one photon, for one channel of the PDU in LN, measured for \SI{7}{\V} over voltage. 
    }
    \label{fig:ivandwf}
    \end{figure}

Several pre-production PDUs have been tested both at ambient temperature and in liquid nitrogen to evaluate their performance against the requirements set by the Collaboration and to assess the variability between different units as well as the temporal stability of individual units. The first measurement performed on a PDU is the I-V characteristic. As SiPMs operate in Geiger mode, their performance is strongly dependent upon the excess voltage applied above their breakdown point, which can be measured from the ankle in the I-V characteristic of the device. Such measurement for a pre-production PDU is reported in Figure~\ref{fig:ivandwf}a, the breakdown voltage being \SI{27.1(1)}{\V} in LN and \SI{32.8(1)}{V} at ambient temperature. The excess voltage above breakdown is the over voltage (o.v.). 

To characterize the PDU in LN, the SiPMs are biased with the nominal over voltage value of \SI{7}{\V} o.v. and exposed to a ps-pulsed laser. %
The average waveform corresponding to the detection of a single photon (one photoelectron) is shown in Figure~\ref{fig:ivandwf}b, fitted with a template function, sum of two exponential components (rise and decay). The measured rise and decay times are \SI{85(5)}{\ns} and \SI{345(5)}{\ns} respectively, with a signal amplitude of approximately \SI{12}{mV}.

\begin{figure}[t]
    \centering
        \includegraphics[height=130pt]{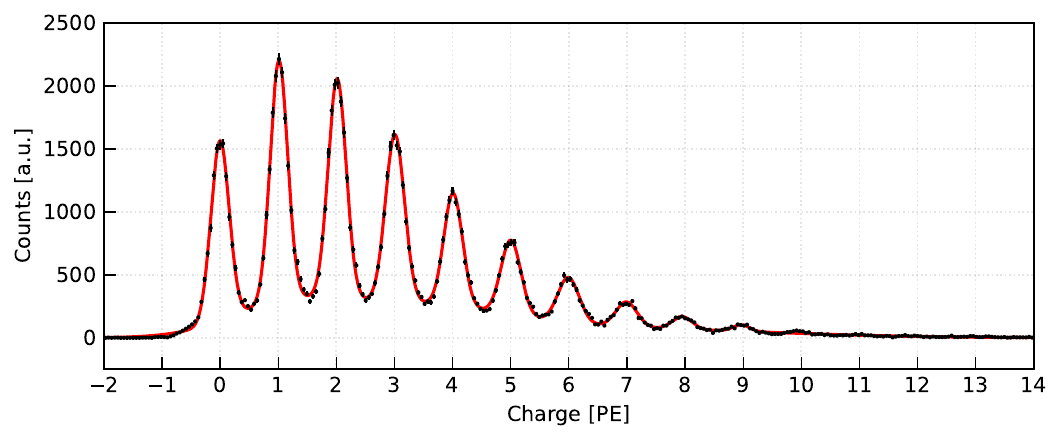}

    \caption{Charge spectrum obtained from one channel of the PDU in LN illuminated with a short-pulsed laser, at \SI{7}{\V} over voltage. Each peak corresponds to a given number of photons, starting from zero with the leftmost peak (pedestal). Together with data points (black) a multi-Gaussian fit is plotted (red).}
    \label{fig:spectrum}
    \end{figure}

Signals obtained from the pulsed laser are integrated to measure the response charge, as plotted in Figure~\ref{fig:spectrum} for a single channel. The charge spectrum exhibits the finger-shaped curve typical of SiPM devices, well described by a sum of multiple Gaussian distributions.%
The \(\sigma/\mu\) ratio for the first photon peak in the charge spectrum is approximately $13\%$, with the dominant contribution being from electronic noise (the precise value depends upon the integration region). From the event distribution, the average number of correlated pulses can be extracted from a compound Poissonian fit~\cite{Vinogradov:2009xxx} and its average value is of \SI{43(3)}{\%} (at \SI{7}{V} o.v.) i.e. for each primary photon an average of 0.43 additional pulses are generated due to cross-talks and afterpulses. 
An important performance metric for a photon detector is the signal-to-noise ratio (SNR). For the PDU at \SI{7}{V} overvoltage, the raw-data SNR is $\sim7$. With a matched filter, it exceeds $10$ -- well within specifications.

All of the reported quantities have been measured as a function of time over the course of several months in stable conditions, and they generally exhibit stability at the level of \SI{1}{\%_{rms}} or better. In particular, the charge measured from the detection of a single photon is stable below \(0.5\%_\text{rms}\), and a similar result is obtain for the amplitude.

\section{Summary}
Cryogenic SiPM arrays are emerging as a key technology in large-scale noble liquid detectors for dark matter searches and related low-background applications. Within the Global Argon Dark Matter Collaboration, extensive R\&D has yielded a robust photosensor design based on a modular approach, each PDU offering four large-area (\SI{100}{\cm^2}) channels. These units are currently being mass-produced at the Nuova Officina Assergi facility and tested at the INFN Naples Photosensor Test Facility, as well as other collaboration sites for the veto PDUs. 

Overall, the PDUs demonstrate high single-photon resolution, stable operation, and adequate noise performance at cryogenic temperatures. More than 600 PDUs will be employed in DS-20k, covering over \SI{27}{\m^2} of total photo-sensitive area in the TPC and veto systems. This effort marks a significant step forward in the use of cryogenic SiPMs for next-generation noble-liquid detectors.

\bibliographystyle{JHEP}
\bibliography{biblio.bib}
\end{document}